\documentclass[12pt]{iopart}

\usepackage{braket}
\usepackage{graphicx}

\begin{document}

\title{Efficient quantum computing with weak measurements}

\author{A.~P.~Lund}
\address{Centre for Quantum Computer and Communication Technology,
Centre for Quantum Dynamics, Griffith University, Nathan Queensland
4111, Australia}
\ead{a.lund@griffith.edu.au}

\begin{abstract}
Projective measurements with high quantum efficiency is often assumed
to be required for efficient circuit based quantum computing. We argue
that this is not the case and show that this fact has actually be
known previously though not deeply explored. We examine this issue by
giving an example of how to perform the quantum ordering finding
algorithm efficiently using non-local weak measurements given that the
measurements used are of bounded weakness and some fixed but arbitrary
probability of success less than unity is required. We also show that
it is possible to perform the same computation with only local weak
measurements but this must necessarily introduce an exponential
overhead.
\end{abstract}



\maketitle

\section{Introduction}

The work of DiVincenzo~\cite{divincenzo2000} states explicit
requirements for scalable circuit based quantum computing. Given the
current state of art, meeting these requirements in even moderately
sized systems is technologically challenging (\cite{ladd10} and
references therein). With some more modern implementations the
criteria can be difficult to apply, but some reinterpreted set of
criteria will apply for any particular implementation~\cite{ladd10}.
DiVincenzo's requirements consist of five criteria: well defined
scalable qubits, the ability to prepare fiducial states, near perfect
(below fault tolerant threshold) unitary evolution, access to a
universal set of unitary evolutions and near perfect quantum
measurement.

There exists an assumption that the measurement criteria requires
strong projective measurements with near unit quantum efficiency to
achieve the efficiency possible in quantum
computing~\cite{strong-meas}. This may seem reasonable given that
proposed quantum algorithms which are efficient compared to the best
known classical algorithms are presented with measurements in the
basis of the eigenstates of Hermitian operators. Furthermore, models
of quantum computing such as cluster state quantum
computing~\cite{rass01} rely on strong measurements to perform the
required state transformations for universal computation. However, as
DiVincenzo mentions~\cite{divincenzo2000}, this is not a strict
requirement and one can make trade-offs between conditions to achieve
scalable quantum computing. The important issue is if when making a
trade-off that algorithmic efficiency is not lost.

This work is motivated by this brief observation of DiVincenzo to
explicitly show a non-trivial example of an efficient quantum
algorithm which involves non-ideal, and in particular, weak quantum
measurements~\cite{aav}. As a result, we hope to demonstrate in theory
that when building a demonstration quantum computer based on the
circuit model, strong projective measurement for read-out in the
computational basis is not absolutely necessary. This is an important
consideration when constructing small to medium scale quantum
computers as it allows for an extra degree of freedom which can assist
in the the design of algorithms matched to the strengths of the
particular architecture used.

In this work we will consider working in an architecture that is
constrained in such a way as the interaction strengths for the readout
measurements will only vary over a very small range and the time taken
for the measurement is limited to small values to minimise the effects
of decoherence. Within this constraint there has been some work to
speed-up the measurement process by adapting parameters as the
measurement proceeds~\cite{combes06}. Here we will consider working in
a non-adaptive regime and allow for arbitrary small (but bounded)
measurement strengths. As the information gained in each measurement
is small the results from any algorithms must necessary be formed by
processing over an ensemble. The situation we will consider is
distinct from the situation found for bulk ensemble nuclear magnetic
resonance quantum computing~\cite{braunstein99} as we will still
require the preparation of pure quantum states before the computation.

Our paper is ordered as follows: In the first section we will describe
weak measurements following the standard presentation given in recent
literature. Then we will then review a specific type of weak
measurement on qubits which differs slightly to the standard
presentation but will be useful for our purposes. In the second
section we will describe how to use this weak measurements in quantum
computing and give two specific examples of algorithms which may
utilise such measurements. The two examples will be of the
satisfiability and order finding algorithms which we will show lead to
a respective inefficient and efficient use of weak measurements in
quantum algorithms. In the penultimate section we will discuss the
potential use of fault tolerant constructions within this model and
the how using local weak measurements generally results in an
inefficient overhead. Finally we will conclude our results.

\section{Weak Measurements}
\label{weak-meas-section}

Aharonov, Albert and Vaidman (AAV)~\cite{aav} shows how one can make a
"weak" measurement of an observable $A$ in which any single
measurement outcome from the apparatus has very little information
about the value of $A$ and is hence very noisy. However, averaging
over a large enough ensemble this noise can be removed and averages of
$A$ can be recovered. It is possible to construct the measurement so
that the lower the information gathered about $A$ the less the system
is disturbed. Quantum mechanics allows this disturbance to go to zero
as the information obtained for $A$ goes to zero~\cite{ozawa}.
However, as the measurement becomes weaker larger ensembles are
required to mitigate the effects of the noise and maintain a desired
precision for the average of $A$.

AAV consider a measurement model with a system Hilbert space and a
separate apparatus Hilbert space which describes the measuring
apparatus. The apparatus space is assumed to have the same structure
as a harmonic oscillator and the observable $X$ will represent the
measurement outcomes and $P_x$ will be the generator of infinitesimal
translations in $X$. The apparatus is also assumed to be in an initial
state which is Gaussian and separable from the system. The system and
apparatus are coupled by a Hamiltonian $H = g A \otimes P_x$ where $A$
is the observable that we wish to weakly measure and $g$ is a scalar
value which will be a factor in determining the strength of the
measurement. The observable $A$ can be any observable on the system. A
system which is strongly isolated will have small values for the
coupling constants in the Hamiltonian.

In the Heisenberg picture, the apparatus observable $X$ evolves to $X
+ gtA$ where $t$ is the interaction time for the coupling between the
system and apparatus. Knowing the strength and duration of the
coupling and the initial state of the apparatus gives sufficient
information for the statistics of $A$ to be calculated from the
measurement results from the apparatus alone irrespective of the
strengths of the interaction. However, weaker measurements will
require more measurements if some bound on the uncertainty in the
statistical estimators is required to be achieved.

\subsection{Projector probability observables}

Projection operators are valid observables. The expectation value of
such a projector observable is the modulus squared length of the
component of the state within the subspace of the projector. In other
words, if the projector is constructed from the space spanned by
eigenstates of an observable with particular eigenvalue, then the
expectation value of the projector is the same as the probability that
a strong measurement of the observable would result in that eigenvalue
had it been made on the same ensemble. This idea of projectors as
probability operators follows naturally from the generalised theory of
quantum measurement.

If one can make a weak measurement of this projector observable then
it is possible to obtain this probability without actually having to
actually perform the strong measurement of the underlying observable
or greatly disturbing the system.

Finding a system with a Hamiltonian of the right form for a projector
observable might be difficult, but one can use the quantum computing
circuit model to construct a device which does with the system and
apparatus both qubits~\cite{pryde05,ralph06,lund10}. This construction
is not the same as that considered in AAV, but of the same flavour. We
will now describe this construction of a qubit weak measurement of a
projector observable.

\subsection{Single-qubit measurement model}

\begin{figure}
\centerline{\includegraphics[width=50mm]{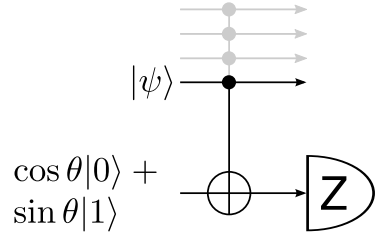}}
\caption{\label{measurement-qubit}Model of a qubit based weak
measurement. An apparatus (the lowest shown qubit) is prepared in the
state $\cos \theta \ket{0} + \sin \theta \ket{1}$ and measured in the
$Z$ basis after interacting with the system (the upper qubits)
prepared in an arbitrary state $\ket{\psi}$. The black lines shows the
case of a single qubit weak measurement in which the interaction
between the system and meter is of the form of a singly controlled not
gate. The black and grey lines combined show the general many qubit
case of a multi-controlled not gate. The weakness of the measurement
is determined by the parameter $\theta$ where for $\cos \theta = 1$
the measurement is strong and equivalent to a projective measurement
and where $\cos \theta = 1/\sqrt{2}$ the measurement is completely
turned off. See text for details.}
\end{figure}

Consider a measurement with the system and apparatus Hilbert spaces
both a single qubit. The system is assumed to be prepared in an
arbitrary state $\ket{\Psi}$ and the apparatus is prepared in a pure
state $\cos \theta \ket{0} + \sin \theta \ket{1}$ uncorrelated with
the system (i.e. a separable state). Instead of specifying a
Hamiltonian we will specify the coupling of the system and apparatus
by a unitary gate, in particular the controlled NOT (CNOT) operation.
Finally the apparatus will be observed with the $Z = \ket{0}\bra{0} -
\ket{1}\bra{1}$ observable. This configuration is depicted in black in
Figure~\ref{measurement-qubit}.

If the $Z$ measurement of the apparatus is propagated back through the
CNOT (c.f Heisenberg picture) then the final measurement of the
apparatus is equivalent to a measurement of $Z \otimes Z$ on the
system and apparatus before the interaction. In other words, the
measurement is equivalent to a measurement of the parity subspace on
the combined input state. If the system state is written $\ket{\Psi} =
\alpha \ket{0} + \beta \ket{1}$ then the probabilities for the two
measurement results will be
\begin{eqnarray}
P(+) & = & |\alpha|^2 \cos^2 \theta + |\beta|^2 \sin^2 \theta \\
P(-) & = & |\alpha|^2 \sin^2 \theta + |\beta|^2 \cos^2 \theta.
\end{eqnarray}
From these equations it is possible to see that for $\theta = 0$, the
measurement output will be equivalent to a projective $Z$ measurement
on the system. For $\theta = \pi/4$ then the output will give either
result with equal probability independent of the system state. It is
possible to show that when $\theta = \pi/4$ the state of the system is
undisturbed. This is unlike the AAV model in which the strength of the
interaction is tuned not only from the initial meter state, but by the
strength of the interaction in the Hamiltonian and the interaction
time. Here the full range of possible measurement strengths are
achieved by tuning the initial meter state. However, one can think of
this model as a $Z$ measurement on the system of the AAV type.

The average value for $Z$ can be found from the expectation value
of the function $\tilde{Z}$ defined by
\begin{eqnarray}
\tilde{Z}(x = +) & = & \frac{1}{\cos^2 \theta - \sin^2 \theta} \\
\tilde{Z}(x = -) & = & -\frac{1}{\cos^2 \theta - \sin^2 \theta}
\end{eqnarray}
where $x$ is the meter measurement result. This function is well
defined for $\theta \in [0,\pi/4)$. The variance of this function on a
single measurement is given by
\begin{equation}
\frac{1}{(\cos^2 \theta - \sin^2 \theta)^2} 
- (|\alpha|^2 - |\beta|^2)^2.
\end{equation}
The variance can be understood as having a contribution of $(\cos^2
\theta - \sin^2 \theta)^{-2} - 1$ from the variance due to the
weakness of the measurement which can be infinitely large and $1 -
(|\alpha|^2 - |\beta|^2)^2$ from the variance of the system state
which is at most $1$. For weak measurements the variance in the output
is dominated by the variance due to the weakness of the measurement.
This statement can be taken as a quantitative definition of
measurement weakness.

It is possible to make a measurement of the expectation value of the
projector onto the $+1$ subspace of the $Z$ operator by the same
apparatus but calculating the expectation value of the function
\begin{eqnarray}
\tilde{\Pi}_Z(x = +)&=&\frac{1}{2}+\frac{1}{2(\cos^2 \theta - \sin^2 \theta)} \\
\tilde{\Pi}_Z(x = -)&=&\frac{1}{2}-\frac{1}{2(\cos^2 \theta - \sin^2 \theta)}
\end{eqnarray}
which has a mean of $|\alpha|^2$ for all theta except $\pi/4$ and a
variance of
\begin{equation}
\frac{1}{4(\cos^2 \theta - \sin^2 \theta)^2} -
 \frac{(|\alpha|^2 - |\beta|^2)^2}{4}.
\end{equation}
A similar analysis of the contributions to the variance can be made as
above.

\subsection{Multi-qubit measurements}

It is possible to extend this construction to build a larger class of
weak measurement of projectors using multiply controlled NOT gates.
This configuration is depicted in the combined black and grey
schematic in Figure~\ref{measurement-qubit}. Multiply controlled
NOT gates can be built efficiently using $O(n^2)$ singly controlled
gates and local unitaries~\cite{mikeandike}. A measurement of $Z$ on
the meter after the interaction is equivalent to a measurement of the
operator
\begin{eqnarray}
\hat{P}_\bot \otimes \ket{0} \bra{0} +
 \ket{111 \cdots}\bra{111 \cdots} \otimes \ket{1} \bra{1} - \nonumber \\
 \hat{P}_\bot \otimes \ket{1} \bra{1} -
 \ket{111 \cdots}\bra{111 \cdots} \otimes \ket{0} \bra{0}
\end{eqnarray}
on the system and meter Hilbert spaces before the interaction where
$\hat{P}_\bot$ is the projector onto the subspace which is the
complement of the all ones subspace (i.e. the subspace which is
spanned by all qubit basis states except $\ket{1111\cdots1}$).

If the apparatus is prepared the as in the case with a single control
and the system is in the state $\ket{\psi}$ then the probability of
the two outcomes of the apparatus measurement are
\begin{eqnarray}
P(+) & = & \bra{\psi} \hat{P}_\bot \ket{\psi} \cos^2 \theta +
 |\braket{111 \cdots | \psi}|^2 \sin^2 \theta \\
P(-) & = & \bra{\psi} \hat{P}_\bot \ket{\psi} \sin^2 \theta +
 |\braket{111 \cdots | \psi}|^2 \cos^2 \theta.
\end{eqnarray}
This distribution is the same as with the singly controlled CNOT gate
but with the probabilities for the qubit in the system being in the
one state replaced by the expectation values of the projectors onto
the space with all ones. The mean and variances as calculated above
also follow this replacement of variables. Therefore the nature of the
statistics do not change as the input size of the system Hilbert space
increases.

\subsection{Measuring probabilities in the computational basis}

This model can also be used to measure the expectation value of
projectors onto any one dimensional subspace generated by a particular
computational basis state by placing $X = \ket{0}\bra{1} +
\ket{1}\bra{0}$ gates before the measurement to transform the desired
subspace into the all ones subspace.

The value of the probability can be read out from the data collected
at the meter by calculating the expectation value for the estimator of
the average of the projection operator given above. Using the
assumption of a weak measurement and large sample sizes, we can
apply the central limit theorem to the estimator for the probability
to calculate the uncertainty in the estimate of the expectation value.
With some fixed error probability $\epsilon$ the estimate confidence
interval is symmetric around the mean value and has width
\begin{equation}
2 \sqrt{2} \frac{\sigma}{\sqrt{M}} \mathrm{erf}^{-1} \left(1 - \epsilon \right) \leq
\frac{\mathrm{erf}^{-1} \left(1 - \epsilon \right)}{\sqrt{2 M}(\cos^2 \theta - \sin^2 \theta)} 
\end{equation}
where $\sigma$ is the standard deviation of the measurement results,
$M$ is the number of measurements made and $\mathrm{erf}$ is the
standard error function.

Measuring the projectors spanned by multiple computational basis
states can be simplified for some particular combinations of states.
If the states contain all combinations of particular qubits with all
other qubits constant, then the qubits which vary can simply be not
measured. However, if even a single qubit combination is missing then
each combination must be measured separately.

\section{Algorithms with weak measurements}

In this section we are going to describe quantum computing algorithms
in terms of the expectation values and decision problems but analyse
the complexity by restricting ourselves to the qubit weak measurement
just described.

\subsection{Algorithmic complexity}

It is assumed that there is some (presumably small) fixed error
tolerance allowed for the computation as a whole. For algorithms
utilising the weak qubit measurement readout just described, the
temporal computational complexity is then determined by how many
repetitions are required to achieve this error value. If under these
conditions the quantum algorithm has a polynomial temporal complexity
it is in the {\bf BQP} complexity class (the class of practical
quantum problems).

We are going to assume that the strength of all measurements is well
known and greater than some fixed constant value. Hence a worst case
value is known for the uncertainty in the output measurement
statistics and we will assume this worst case value is the actual
estimate for the uncertainty. We are also going to assume that sample
sizes are large enough that the central limit theorem applies. We are
not going to be dealing with any distributions in which the central
limit theorem is not valid. These assumptions combined allows the
variance of the sample mean to be computed and hence a signal to noise
ratio involving the estimated mean and the worse case standard
deviation can be used to infer the maximum probability of error
inherent in the computation.

\subsection{Satisfiability with expectation values}

The satisfiability problem is defined as identifying if a logical
statement described by a Boolean function $f$ has a set of inputs
which result in the function evaluating to true. If the function
represents a conjunction of logical statements (the inputs), then the
statement (the output) is said to be satisfied by the particular
combination of truth values used to achieve this output. This problem
is in the class of decision problems.

Cory~et~al.~\cite{cory97} construct a quantum algorithm for solving
the satisfiability problem. In their paper they assumed that the
standard model of quantum computing is enhanced by special
measurements which can extract expectation values of observables for a
single instance of a quantum state in an error and noise free way.
Their work was motivated by the Nuclear Magnetic Resonance quantum
computing model so this type of measurement involving ensemble
averages is a natural consequence of the output signals from that type
of computation. They then show that given an equal superposition of
all possible logical inputs to the function, of which there are $2^n$
possibilities, a unitary which implements $f$ can be built efficiently
and evaluates all of these possibilities coherently in superposition.
The unitary is built so that the output value of the function is
written onto another qubit register which is zero if the input does
not satisfy the statement and one if it does. The expectation value of
the output register is then obtained using the special measurement
which they added as described above. If the expectation value is
non-zero then the logical statement is satisfiable. Though this does
not say which input will satisfy the function, it does show that such
a satisfying input exists. Provided one has this enhancement which
allows for the immediate extraction of expectation values this is a
method of solving an NP-complete problem (i.e. satisfiability)
deterministically in polynomial time.

This result is only possible when neglecting the noise in the output
of such a measurement. If one requires this measurement to be a
standard quantum measurement rather than the special one used, then
the complexity will change as more measurements are required to
counter the effects of the noise. Consider the possible case of where
only a single particular input satisfies the function (as is possible
for any size satisfiability problem). The measurement then needs to
distinguish between the two cases of being unsatisfiable and the
output register in the state $\ket{0}$ and the case of having a single
satisfying input and the output register is in the state
$(1-2^{-n})\ket{0}\bra{0} + 2^{-n}\ket{1}\bra{1}$. The probability to
be estimated is then of size $2^{-n}$ and in the large $M$ limit the
noise is $2^{-n/2}/\sqrt{M}$ and hence the signal to noise ratio
decreases exponentially in the size of the input. This overcomes the
apparent speed-up offered by the enhanced model of quantum computing
considered as the sample size needed to achieve a particular
probability of error in the decision problem will increase
exponentially.

As we show next, not all useful ensemble averages from the output of
quantum computations necessarily have this problem.

\subsection{Order finding with expectation values}

The order finding problem is a critical part of the quantum prime
factorisation algorithm~\cite{shor}. The definition of the order
finding problem is given positive integers $N$ and $x < N$ find the
least positive integer $r$ such that $x^r = 1 (mod N)$. This problem
is an instance of the hidden subgroup problem which is a more general
class of problems~\cite{mikeandike}. The problem of factoring integers
can be reduced to this problem~\cite{mikeandike}. Currently, the best
known classical algorithms have exponential complexity.

The quantum order finding algorithm utilises a quantum modular
exponentiation operation defined by the unitary
\begin{equation}
U_{x,N} = \ket{y} = \ket{xy (mod N)}
\end{equation}
which can be done efficiently using $O(n^3)$ gates where n is the
number of bits needed to represent integers up to $N$. The eigenvalues
of this unitary are $exp(2\pi i s / r)$ where $r$ is the order of $N$
and $s$ is an integer satisfying $0 \leq s < r$ which labels each of
the eigenstates. Therefore performing quantum phase estimation on an
eigenstate of the modular exponentiation operator is a method of
finding information about the order of $x$~\cite{mikeandike}. However
preparing the eigenstate would require that the order of the integer
of interest be known already. Therefore, a superposition state of all
possible eigenstates is used. This state happens to be equal to a
state that is the representation of the multiplicative identity in the
computational basis. Therefore, the output of the order finding
algorithm is a phase $\phi = s/r$ where $r$ is the order that we
desire and $s$ is equally distributed amongst the allowed values. The
continued fractions expansion of $\phi$ allows for the computation of
values for $r$. However, if $s$ and $r$ share a factor, then this
method will give the value of $r$ with this factor divided out. This
is then not the order that was desired but a factor of the order.

\begin{figure}
\centerline{\includegraphics[width=8cm]{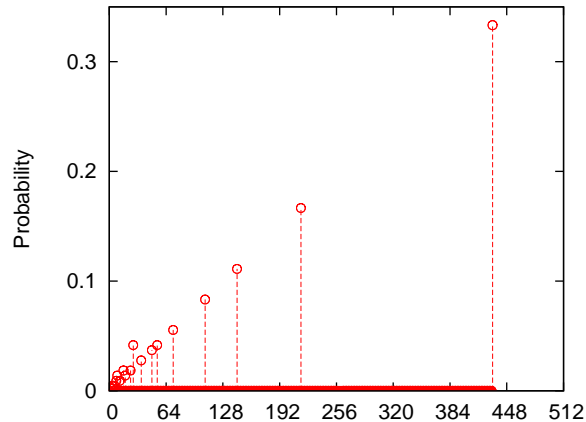}}
\caption{\label{distribution}An example distribution of output values
for the denominator register when the order is known to be $432$. If
the order is prime then all of the probability is concentrated at the
highest value.}
\end{figure}

For a randomly selected value of $s$ the probability that it is prime
for large values of $N$ is at least and will asymptotically approach
\begin{equation}
\label{prime-probability}
\frac{1}{2 \log N} \geq \frac{1}{n (2\log 2)}.
\end{equation} 
This guarantees that there will be some probability that the
correct answer is contained within the output and that this
probability drops as $\Omega(n^{-1})$ with the size of the problem. An
example of the distribution for values of $r$ read out from the
continued fractions algorithm is shown in Figure~\ref{distribution}.

\begin{figure}
\centerline{\includegraphics[width=85mm]{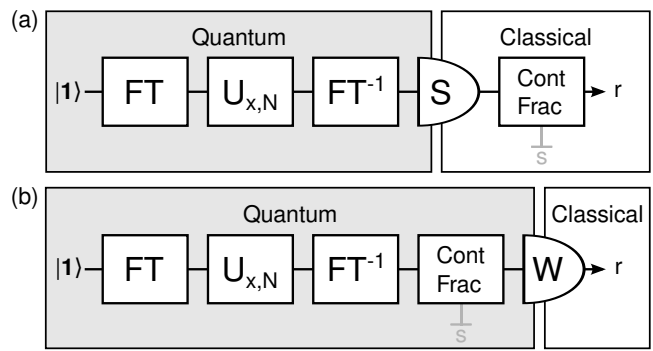}}
\caption{\label{algo}Schematic diagrams showing the difference between
the strong (a) and weak (b) measurement versions of the ordering
finding algorithm. FT represents the Fourier transform unitary and the
unitary $U_{x,N}$ is defined in the text. S and W represent strong and
weak measurement read-out respectively. The continued fraction
algorithm as utilised for this order finding algorithm has two
outputs, the numerator represented by $s$ and whose information is not
utilised, and the denominator register represented by $r$. They key
distinction is that in this presentation the continued fractions
algorithm is performed quantum mechanically in the weak measurement
version.}
\end{figure}

We are now in a position to describe the quantum order finding
algorithm using only weak measurements. The procedure that will be
described here is also shown schematically in Figure~\ref{algo}.
First, build the order finding algorithm as done for projective
readout measurement, but do not measure the register containing the
phase $\phi=s/r$ result. This requires no projective measurements only
good state preparation and precise unitary evolution. Second,
implement the continued fractions algorithm and calculate the rational
convergent on the $s/r$ register in the computational basis quantum
mechanically using a construction based upon universal reversible
gates~\cite{mikeandike}. This construction requires no measurement or
feed-forward, but does require a multi-qubit conditional unitaries.
This shifts $O(n^3)$ classical gates to quantum gates and represents
part of the overhead to this method. Tracing over the numerator, the
reduced density operator for the convergent's denominator register
will be
\begin{equation}
p \ket{\Psi}\bra{\Psi} + (1-p) \hat{\rho}_e
\end{equation}
where $\ket{\Psi}$ is the state of the denominator register
representing $r$ (the result). $\hat{\rho}_e$ is a density operator
orthogonal to $\ket{\Psi}$ representing those terms when the numerator
had a common factor with the denominator. The standard procedure for
the readout is to make a strong projective measurement on this state
in the computational basis to read out a result, test to see if it is
a solution and repeat the algorithm (possibly modified) if the order
if found not to be correct but a factor of the order. Here, we wish to
only use weak measurements to extract the answer. It is clear that the
largest value of any component from the denominator register will be
the order we are seeking and not merely a factor of the order.
Therefore we propose to extract the register state with the largest
numerical value through a bi-sectional search on properties of
denominator register using ensemble averages.

\begin{figure*}
\centerline{\includegraphics{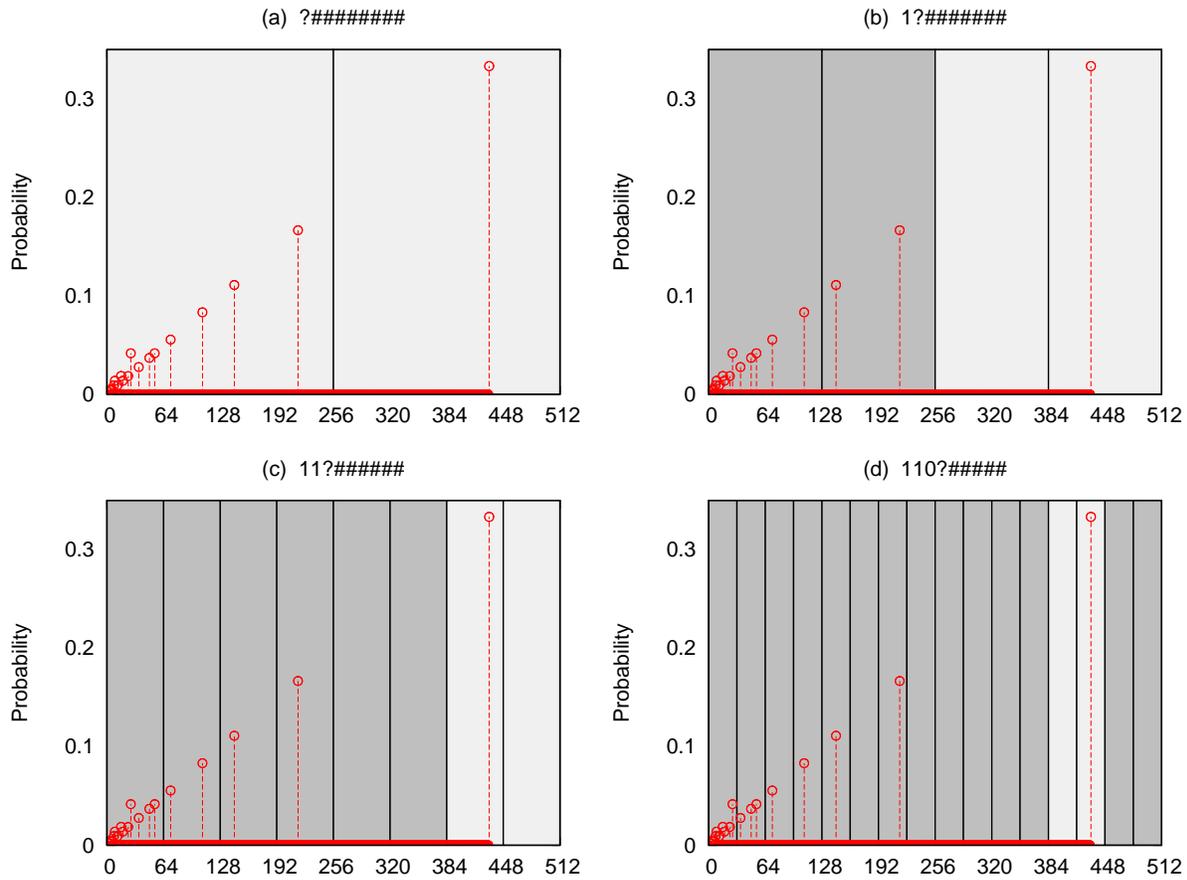}}
\caption{\label{bisection}An illustration of the steps involved for
the bi-sectional search for the read out of the largest value of the
register with non-zero probability. The distribution chosen is the
same as Figure~\ref{distribution}, hence the answer read out should be
a binary representation of $432$ ($110110000$). Initially the most
significant bit is weakly measured, the remaining qubits are not
measured and their values are not yet known. This state of knowledge
is shown in the title of the plot of part (a) with $\#$ representing
unknown information that is not to be measured and $?$ representing
unknown information that is currently being measured. The weak qubit
measurement described in the text is used to determined if there is
any probability of the most significant bit value being one. The
measurement is equivalent to determining if there is any probability
on the right hand side of the plot as shown by the central dividing
line. If there is any probability that the state is one (as is the
case here), then the measurement proceeds to determining if there is
any probability in the output within the $\ket{11}$ subspace as
depicted in part (b). If there was, then the measurement proceeds to
determining if there is any probability in the $\ket{01}$ state. This
repeats until all qubits are measured and the final read out value
will be the largest non-zero probability value within the qubit
register. If there are $n$ qubits then this procedure has $n$ steps.}
\end{figure*}

The bi-sectional search proceeds by a series of decision problems. The
problems form the answer bit by bit generating the largest value with
non-zero probability from the most significant bit to the least
significant bit. These probabilities are extracted by taking
expectation values of carefully selected projectors as we will now
detail. Consider the projector onto the space containing a logical one
state for the most significant qubit of the denominator register and
all other registers allowed any value. This projector is
\begin{equation}
\hat{\Pi}_{1...} = \frac{1}{2}(Z - I) \otimes I \otimes I \otimes
\ldots.
\end{equation}
If the average of this projector was non-zero, then it is known that
the largest numerical value in the computational basis for the
denominator register state have its most significant bit as a 1. If
the average is zero then the largest value must have a zero for the
most significant bit. This procedure is this repeated for the next
most significant bit using the appropriate projector adapted from
the previous result. For example, if the most significant bit was a
zero, then the next measurement to be made would be
\begin{equation}
\hat{\Pi}_{01...} =
 \frac{1}{2}(Z + I) \otimes \frac{1}{2}(Z - I) \otimes I \otimes
\ldots
\end{equation}
or if it was a one then the next measurement is
\begin{equation}
\hat{\Pi}_{11...} =
 \frac{1}{2}(Z - I) \otimes \frac{1}{2}(Z - I) \otimes I \otimes
\ldots.
\end{equation}
This continues for each bit and when all have been read out the
largest value with non-zero probability for the register is known. At
each step the projector representing the space containing the answer
has it's dimension halved however the size of the expectation value is
bounded by the probability given in Eq.~\ref{prime-probability}. An
illustration of how the bi-sectional search works is shown in
Figure~\ref{bisection}.

To achieve an overall algorithmic error probability less than
$\epsilon$, the error probabilities for each bit readout measurement
must be less than $\epsilon/n$. This is because if the probability of
failure for one run is $\epsilon^\prime$ then the overall probability
of failure is $1-(1-\epsilon^\prime)^n < n \epsilon^\prime$ and hence
$n \epsilon^\prime < \epsilon$. If we invoke the central limit theorem
as foreshadowed above and assume that the estimator for the mean is
normally distributed with a variance of $\sigma_0^2 / n$ with
$\sigma_0$ being the variance in a single outcome and we are deciding
between two means of $s_0$ and $s_1$ (which we will call the signal),
then we define $\mathrm{SNR}_0 = |s_1 - s_0|/\sigma_0$. Here we are
assuming that the variance from the two distributions is equal. We can
do this by choosing a worse case variance as described above. Taking a
threshold half way between the two signal values, the probability of
making an error is
\begin{equation}
P_{\mathrm{error}} = 1 - \Phi\left(\frac{\mathrm{SNR}_0 \sqrt{M}}{2\sqrt{2}}\right)
\end{equation}
where $\Phi$ is the cumulative distribution function of the standard
normal distribution. We require this probability to be less than
$\epsilon^\prime = \epsilon/n$. The number of samples required to meet
the error budget must therefore satisfy
\begin{equation}
\label{number}
M = \left[\frac{2\sqrt{2}}{\mathrm{SNR}_0} \mathrm{erf}^{-1}
 \left(1 - \frac{2 \epsilon}{n}\right)\right]^2
\end{equation}
which scales poly-logarithmically in $n$ for the bi-sectional search
alogrithm. To prove this scaling, rearranging this expression gives
\begin{equation}
\mathrm{erf}\left(\frac{\mathrm{SNR}_0 \sqrt{M}}{2\sqrt{2}}\right)
 = 1 - \frac{2 \epsilon}{n},
\end{equation}
which is equivalent to
\begin{equation}
\frac{\epsilon}{n} = \frac{1}{\sqrt{\pi}} \int_{\mathrm{SNR}_0
\sqrt{N}/2\sqrt{2}}^\infty e^{-t^2} \mathrm{d}t.
\end{equation}
For $M$ sufficiently large (in particular $\mathrm{SNR}_0 \sqrt{M} > 2 \sqrt{2}$)
\begin{eqnarray}
\int_{\mathrm{SNR}_0 \sqrt{M}/2\sqrt{2}}^\infty e^{-t^2} \mathrm{d}t &<&
\int_{\mathrm{SNR}_0 \sqrt{M}/2\sqrt{2}}^\infty e^{-t} \mathrm{d}t \nonumber \\
&=& e^{-\mathrm{SNR}_0 \sqrt{M}/2\sqrt{2}}.
\end{eqnarray}
Therefore
\begin{equation}
n > \epsilon \sqrt{\pi} e^{\mathrm{SNR}_0 \sqrt{M}/2\sqrt{2}}
\end{equation}
or rearranging
\begin{equation}
\sqrt{M} < \frac{2\sqrt{2}}{\mathrm{SNR}_0} ( \log n - \log \epsilon\sqrt{\pi} )
\end{equation}
and therefore
\begin{equation}
\left[\frac{2\sqrt{2}}{\mathrm{SNR}_0} \mathrm{erf}^{-1}
 \left(1 - \frac{2 \epsilon}{n}\right)\right]^2 = O(n \log(n)^2)
\end{equation}
where we have used the $\Omega(n^{-1})$ scaling of $\mathrm{SNR}_0$
from equation~\ref{prime-probability}. Therefore the number of total
weak measurement samples needed in the algorithm is
$\Omega(n\log(n)^2)$.

This requirement to make poly-logarithmically extra samples forms
another part to the overhead of this procedure. Furthermore, the
multiplying factor in the scaling will depend on the weakness of the
measurement which may be large for very weak measurements. However,
none of the overheads introduced in this presentation scale
exponentially in the size of the input.

\section{Discussion}

\subsection{Local weak measurements}

Resch and Steinberg~\cite{resch04} have shown that it is possible to
extract non-local weak values from local weak measurements. Therefore,
one might be tempted to measure the multi-qubit expectation value
using local single-qubit measurements instead of the non-local
measurements used here. However this does introduce an inefficient
overhead.

In general, measuring an $n$ qubit output will need to have estimates
of the expectation values for observables of the form $A_1 A_2 \cdots
A_n$. When observing the correlations in the local meter readouts to
estimate this value, the variance of the correlation constructed from
all the meters is
\begin{equation}
var(X_1 X_2 ... X_n)_o =
 \langle (X_1 X_2 ... X_n)^2 \rangle_o - \langle X_1 X_2 ... X_n \rangle_o^2,
\end{equation} 
where $X$ represents the meter observables as per
section~\ref{weak-meas-section} and the subscript $o$ is a reminder
that this description is for measurements at the meter output. If each
meter is initialised separately with a mean of zero and a variance of
$\sigma^2$ then the variance at the meter output in terms of the
inputs becomes
\begin{eqnarray}
var(X_1 X_2 \cdots X_n)_o & = &
 \langle (X_1 + \gamma A_1)^2 (X_2 + \gamma A_2)^2 \cdots (X_n + \gamma A_n)^2 \rangle_i
\nonumber \\ &&- \gamma^{2n} \langle A_1 A_2 ... A_n \rangle_i^2 \\
& = & \sigma^{2n} + 
 \gamma^2 \sigma^{2(n-1)} \sum_{k=0}^{n} \langle A_k^2 \rangle_i  +
\nonumber \\ && \gamma^4 \sigma^{2(n-2)} \sum_{k=0}^{n} \sum_{l=0}^{n}
\langle A_k^2 A_l^2 \rangle_i +
\cdots + \nonumber \\ && \gamma^{2n} var(A_1 A_2 \cdots A_n).
\end{eqnarray}
where we have used the commutativity of the different subsystems to
rearrange terms and statistical independence of the meters and the
meter and system to remove terms. This expression has a scaling of
$O(\sigma^{2n})$ from the first term on the right hand side which is
independent from the actual signal from the system observable.
Therefore the $\mathrm{SNR}_0$ term, instead of being constant as is
the case above, decreases exponentially entirely removing the
efficiency of the algorithm presented.

Observables of the type just mentioned are observed locally in the
standard presentation of quantum computing algorithms using strong
measurements. Clearly there must be a point of transition in the
initial variance of the meter states compared to the measurement
strength where the exponential scaling term from the meter noise does
not play a significant role in the data extraction. This quantity will
be dependent on the observables needed and hence the type of algorithm
being implemented. For example a fault tolerant implementation would
have a point in which the noise scaling reduces as the size of the
observables increased rather than increasing as is the case in this
simple example.

Another possible way to avoid this problem of local weak measurement
introducing an exponential overhead is to break the requirement of
local preparation of initial meter states. If the initial meter state
was correlated then the equality reached above would change
significantly. If the right state is chosen for the observable of
interest then it may be possible to avoid the exponential scaling.

Other work on non-local weak measurements in a completely different
context has also found that estimating non-local correlations is
inefficient and requires large ensembles~\cite{kedem10,broduch09}. So
it appears that for efficient quantum information processing with weak
measurements, non-locality is strictly required. This means that
schemes for extracting conditional expectation values using
informationaly complete but not full strength
measurements~\cite{ralph08} cannot be used to perform efficient
computation.

\subsection{Decoherence times}

One may argue that the weakness of the readout and the length of time
for the output for the algorithm counteract one another. However, this
is not true for the algorithm presented here as the algorithm is
rerun, qubits are reprepared and the unitary evolution is run again,
which removes any of the effects of previous decoherence. Hence the
important time to consider is decoherence over the time taken to
execute all the operations needed to run the algorithm in total just
is the case for strong measurements. With the standard model of weak
measurements (as presented here and in~\cite{aav}) the interaction
time for a weak measurement is much smaller than that for the
corresponding strong measurement and hence could act to reduce the
effects of decoherence.

\subsection{Fault Tolerance}

Fault tolerance encoding, evolution and decoding can still be
performed if the final measurements are not strong measurements. For
the CSS class of quantum codes one can avoid using measurements
completely and still achieve fault tolerance~\cite{benorr}. Doing so
involves some penalty in the fault tolerant threshold, but as shown
recently this penalty is not as great as has been believe
previously~\cite{pazsilva10}.

\subsection{Implications for experimental implementation of quantum
computing}

This result suggests that in the pursuit of preliminary or
proof-of-principle quantum computing experiments that strong isolation
and high fidelity operations are where effort should focus provided
one has the ability to readout data even if very noisy. For the order
finding algorithm presented here, having a weak readout does not harm
the efficiency of quantum computing. Increasing the strength of the
readout clearly has an advantage in the rate at which computation can
occur, but this should not be done to the detriment of the ability for
the data to be preserved within the quantum computer to complete the
computation.

\section{Conclusion}

We have outlined how weak measurements in quantum computing can be
modelled theoretically and modified a quantum algorithm using this
model in such a way that the computational efficiency of performing
the algorithm quantum mechanically is maintained. The requirements on
state preparation and control over the evolution are the same as for
any other model of quantum computation. This may be able to assist in
the technological challenge of building demonstration quantum
computers.

\ack

APL acknowledges the fruitful discussions with H.~M.~Wiseman and
T.~C.~Ralph when preparing this manuscript. This research was
conducted by the Australian Research Council Centre of Excellence for
Quantum Computation and Communication Technology (project number
CE11E0096). This research was also supported by the Griffith
University Postdoctoral Research Fellowship.

\end{document}